\documentclass[floatfix,notitlepage,twocolumn]{revtex4-1}%
\usepackage{graphicx,color}%
\usepackage{amsmath}%
\usepackage{txfonts}
\usepackage{amsfonts}%
\usepackage{amssymb}
\usepackage{bm}
\usepackage[breaklinks]{hyperref}

\usepackage{ulem}
\usepackage[usenames]{xcolor}

\def\s{{\sigma}}

\def\k{{ {\bm k} }}

\def\0{{ {\bm 0} }}

\def\a{{\alpha}}
\def\b{{\beta}}							
\def\g{{\gamma}}

\def\L{{\Lambda}}
\def\r{{ {\bm r} }}
\def\dg{{ \dagger }}
\def\expo{{ {\rm e} }}
\def\rA{{ {\rm A} }}
\def\rB{{ {\rm B} }}
\def\ong{{ \mathrm{\mathring{A}} }}
\allowdisplaybreaks[4]

\begin{document}
\title{ 
Asymmetric modulation of Majorana excitation spectra and nonreciprocal 
thermal transport 
in the Kitaev spin liquid under a staggered magnetic field
}
\author{
Kazuki Nakazawa\thanks{k.nakazawa@aion.t.u-tokyo.ac.jp}, 
Yasuyuki Kato, and 
Yukitoshi Motome
}

\date{\today }

\begin{abstract} 
We present the theoretical results for the
Majorana band structure and thermal transport properties in the Kitaev honeycomb model under a staggered magnetic field in addition to a uniform one. 
The Majorana band structure is asymmetrically modulated in momentum space, and the valley degree of freedom is activated and controlled by the magnitudes and directions of the uniform and staggered magnetic fields; as a result, the Majorana excitation is either gapped or gapless, the latter of which in general has Majorana Fermi surfaces, in contrast to the point nodes in the absence of the fields.  
We show that the asymmetric deformation of the Majorana band leads to a nonreciprocal thermal transport.
The field dependence of the linear thermal conductivity is correlated with the magnitude of the Majorana gap, while that of the nonlinear thermal conductivity is associated with the asymmetry of the Majorana excitation spectra. 
We show the estimates of the thermal currents and discuss that the linear response is experimentally measurable, whereas future improvement is required for the observation of the nonlinear one. 
We also discuss how the thermal currents are modulated by other additional effects of the magnetic field and contributions from conventional magnons, latter of which are expected in heterostructures to realize an internal staggered magnetic field. 
\end{abstract}

\address{
Department of Applied Physics, University of Tokyo, Tokyo 113-8656, Japan
}

\maketitle

\section{Introduction}
\label{sec:Intro}

The quantum spin liquid (QSL) is a spin state in insulating magnets which does not break any symmetry of the systems down to absolute zero temperature~\cite{Balents, SB, KM}.  
This exotic state is expected to be realized when magnetic frustration and quantum fluctuation are in action. 
Among many proposals, the Kitaev honeycomb model has drawn attention since its ground state provides a rare example of the QSL in more than one dimension~\cite{Kitaev2006}. 
In the Kitaev spin liquid, the elementary excitations are described by itinerant Majorana fermions and localized $Z_2$ fluxes as a consequence of the fractionalization of the spin degree of freedom~\cite{Kitaev2006}. 
The quantum entanglement between the fractional excitations would be utilized for decoherence-free topological quantum computing~\cite{Kitaev2003}. 
Since it was pointed out that a class of materials with strong spin-orbit coupling and electron correlation is a good platform for the realization of the Kitaev model~\cite{JK}, extensive investigation has been performed toward the identification of the Kitaev spin liquid and the fractional excitations for the candidate materials, such as $A_2 {\rm Ir O_4}$ $(A={\rm Li \ and \ Na})$ and $\a$-RuCl$_3$~\cite{TTJKN,MN,Winter,Trebst}. 

To capture the evidence of the fractionalization, the effect of a magnetic field has been studied both theoretically and experimentally. 
In theory, the magnetic field opens a gap at the Dirac-like nodal points at the Brillouin zone edges in the excitation spectrum of the itinerant Majorana fermions, and the system becomes a Majorana Chern insulator with a chiral Majorana edge mode in the weak field limit, which leads to the half-quantized thermal Hall effect~\cite{Kitaev2006,NYM}. 
Experimentally, the external magnetic field suppresses the parasitic long-range order stabilized by residual non-Kitaev interactions and may realize the Kitaev spin liquid in the field~\cite{Wolter,Jansa,Banerjee2018,Zheng,Nagai}. 
Indeed, the half-quantized thermal Hall effect was observed in the field-induced quantum paramagnetic region in one of the candidates, $\a$-RuCl$_3$~\cite{Kasahara,Yokoi,Yamashita,Bruin}, which is regarded as strong evidence of the chiral Majorana edge modes in the gapped  topological Majorana state.

Recently, modulations of the Majorana excitation spectrum by the magnetic field have been examined in a wider scope. 
For instance, an asymmetric modulation of the Majorana band structure was discussed as a combined effect of a site-dependent internal magnetic field and a non-Kitaev interaction~\cite{TF2}. 
A similar but different type of asymmetric modulation was discussed by introducing an electric field  in addition to the magnetic field~\cite{CMR}. 
In both cases, the sublattice symmetry is broken, which activates the valley degree of freedom and may bring about the Majorana Fermi surfaces depending on the parameters. 
Such ``Majorana band engineering" would be useful for identifying the fractional excitations in the Kitaev spin liquid, but the systematic study remains yet to be investigated, especially the consequence on the thermal transport which is sensitive to the modulations of the Majorana excitation spectrum. 

In this paper, we perform a systematic study of the Majorana excitation spectrum and the thermal transport in the Kitaev model, by introducing both uniform and staggered magnetic fields. 
Within the perturbation theory in terms of the magnetic fields, we show that the Majorana band structure is modulated in an asymmetric way depending on the magnitudes and directions of the magnetic fields, which allows us to control the valley degree of freedom. 
Furthermore, it becomes gapless with the formation of the Majorana Fermi surfaces in some parameter regions of the magnetic fields. 
For these situations, we calculate the longitudinal thermal transport coefficient by using the Boltzmann transport theory. 
We find that the thermal transport becomes nonreciprocal due to the asymmetric modulation of the Majorana band structure. 
By the comprehensive study while changing the uniform and staggered fields, we show that the linear component of the thermal current probes the Majorana excitation gap, while the nonlinear one is sensitive to the asymmetry of the Majorana band structure.  
We discuss our results with the estimates of the thermal currents expected in real materials. 
We also discuss other additional effects of the magnetic field, including the interaction between the Majorana fermions. Finally, we mention about contributions from the conventional magnons, which would be important in heterostructures with an antiferromagnet to realize an internal staggered magnetic field. 

This paper is organized as follows. 
In Sec.~\ref{sec:Model}, we introduce the model that we use in this paper. 
In Sec.~\ref{sec:Band}, we discuss the Majorana excitation spectra while changing the uniform and staggered magnetic fields, with emphasis on the Majorana excitation gap, the effective Majorana density, and the asymmetry of the Majorana band structure. 
In Sec.~\ref{sec:thermal}, we present the results of the linear and nonlinear thermal transport coefficients for the modulated Majorana bands. 
In Sec.~\ref{sec:disc}, we discuss quantitative estimates of the thermal currents, the effect of the second-order perturbation and the Majorana interaction, and the contributions from the conventional magnons. 
Section~\ref{sec:summary} is devoted to the summary.

\section{Model}
\label{sec:Model}

\begin{figure}[t]
\hspace*{-2mm}
  \includegraphics[width=89mm]{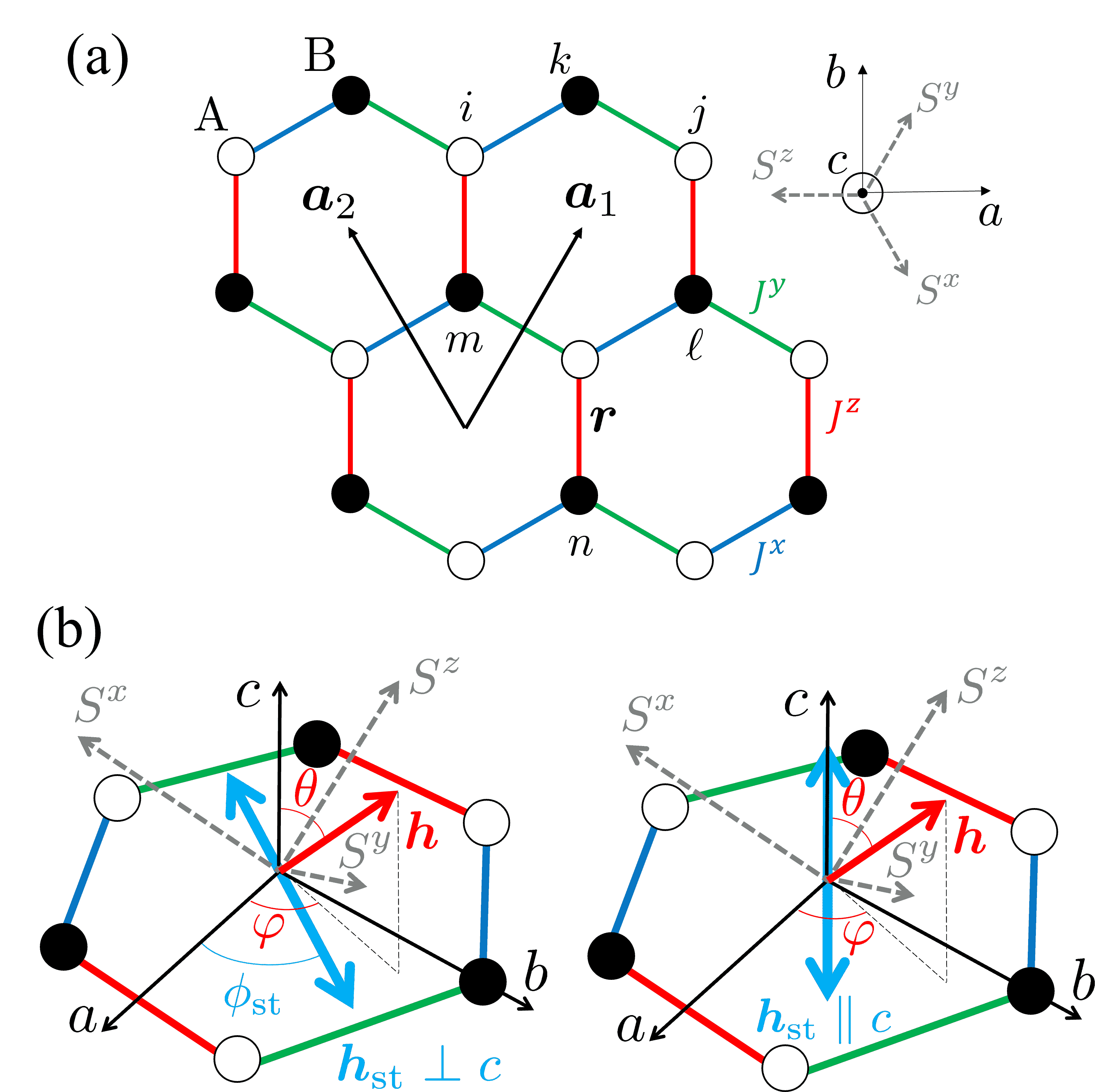}
 \caption{ 
(a) 
Schematic of the Kitaev honeycomb model. 
The white and black circles represent the lattice sites on the A and B sublattices, respectively. 
The blue, green, and red lines represent the $x$, $y$, and $z$ bonds in Eq.~(\ref{eq:H_K}), respectively. 
${\bm a}_1$ and ${\bm a}_2$ denote the primitive translation vectors, $i,j,k$ and $\ell,m,n$ represent examples of the site indices in the summations in Eqs.~(\ref{eq:Hdash}) and (\ref{eq:H4}), respectively, and $\r$ labels the $z$ bond in Eqs.~\eqref{eq:H_K_Maj} and \eqref{eq:Hp}. 
The real-space and spin-space coordinates are shown in the inset. 
A thermal gradient is applied along the $a$ or $b$ direction. 
(b) 
Directions of the uniform component of the magnetic field, ${\bm h}$ (red arrows), and the staggered one ${\bm h}_{\rm st}$ (blue arrows); see Eqs.~(\ref{eq:H_iA}) and (\ref{eq:H_iB}). 
The left panel shows the case with the in-plane $\bm{h}_{\rm st} \perp c$, while the right one is for the out-of-plane $\bm{h}_{\rm st} \parallel c$; $\theta$ and $\varphi$ denote the polar and azimuth angles for $\bm h$, and $\phi_{\rm st}$ is the azimuth angle for the in-plane ${\bm h}_{\rm st}$.  
}
 \label{fig:honeycomb}
\end{figure}

We start from the Kitaev model on the honeycomb lattice under the magnetic field which includes both uniform and staggered components. 
The Hamiltonian is given by ${\cal H} = {\cal H}_{\rm K} + {\cal H}_{\rm Z}$, where 
\begin{align} 
{\cal H}_{\rm K} &= - \sum_{\alpha=x,y,z} \sum_{\langle i,j \rangle_\a} J^\a S_i^\a S_j^\a ,  
\label{eq:H_K}
\\ 
{\cal H}_{\rm Z} &= -\sum_{i} {\bm H}_i \cdot {\bm S}_i . 
\label{eq:H_Z}
\end{align} 
Here, $S_i^\a = \s_i^\a /2$ is the $\a$ component of the spin-1/2 operator at site $i$ defined by the Pauli matrix $\s_i^\a$ ($\alpha=x,y,z$; we set $\hbar = 1$ here and for a while), $J^\a$ is the coupling constant for the Kitaev exchange interaction on the $\a$ bond, and ${\bm H}_i$ denotes the magnetic field at site $i$ given by
\begin{align}
{\bm H}_{i \in {\rm A}} &= {\bm H} - {\bm H}_{\rm st} 
= H \left( {\bm h} - r_{\rm st} {\bm h}_{\rm st} \right) 
\equiv 
H {\bm h}_{\rm A}, 
\label{eq:H_iA}
\\ 
{\bm H}_{i \in {\rm B}} &= {\bm H} + {\bm H}_{\rm st} 
= H \left( {\bm h} + r_{\rm st} {\bm h}_{\rm st} \right)  
\equiv 
H {\bm h}_{\rm B},  
\label{eq:H_iB}
\end{align}  
for site $i$ belonging to the sublattice A and B, respectively (see Fig.~\ref{fig:honeycomb}). 
In Eqs.~\eqref{eq:H_iA} and \eqref{eq:H_iB}, ${\bm H}$ and ${\bm H}_{\rm st}$ denote the uniform and staggered components, respectively; $H=|{\bm H}|$, $|{\bm h}| = |{\bm h}_{\rm st}| = 1$, and $r_{\rm st} = |{\bm H}_{\rm st}|/|{\bm H}|$. 

Following the perturbation theory with respect to the uniform magnetic field up to third order~\cite{Kitaev2006}, 
the Zeeman coupling term ${\cal H}_{\rm Z}$ in Eq.~\eqref{eq:H_Z} can be effectively described by three-spin interactions as 
\begin{align}
{\cal H}' &= - \delta \sum_{\{ i , j , k \}} h_i^x h_j^y h_k^z S_i^x S_j^y S_k^z ,
\label{eq:Hdash} 
\end{align} 
where ${\bm h}_i \equiv {\bm H}_i / H = {\bm h}_{\rm A(B)}$ for $i \in {\rm A(B)}$, and $\delta \sim H^3 / \Delta_{\rm f}^2$ with $\Delta_{\rm f}$ being a (mean) flux excitation energy. 
In Eq.~\eqref{eq:Hdash}, the summation is taken for neighboring three sites $\{ i , j , k \}$ as exemplified in Fig.~\ref{fig:honeycomb}(a). 
By using the Jordan-Wigner transformation~\cite{CH,FZX,CN}, both ${\cal H}_{\rm K}$ and ${\cal H}'$ can be written in terms of the Majorana fermions as 
\begin{align}
{\cal H}_{\rm K} &=
     \frac{i}{4} \sum_{\r} \Bigl[ J^x c_{\r, \rA} c_{\r + {\bm a}_1,\rB}
   + J^y c_{\r,\rA} c_{\r + {\bm a}_2,\rB}
   + J^z c_{\r,\rA} c_{\r,\rB} \Bigr] ,
\label{eq:H_K_Maj}
\\
{\cal H}' &= -\frac{i\delta}{8} \sum_\r 
\Bigl[ 
\tilde{h}_{\rm A}^{yzx} c_{\r - {\bm a}_3,\rA} c_{\r,\rA} 
+
\tilde{h}_{\rm A}^{zxy} c_{\r,\rA} c_{\r - {\bm a}_2,\rA}
+ 
\tilde{h}_{\rm A}^{xyz} c_{\r - {\bm a}_1,\rA} c_{\r,\rA} 
\nonumber \\
&\quad \quad \quad \ 
+ 
\tilde{h}_{\rm B}^{yzx} c_{\r,\rB}  c_{\r - {\bm a}_3,\rB} 
+ 
\tilde{h}_{\rm B}^{zxy} c_{\r  - {\bm a}_2,\rB} c_{\r,\rB} 
+ 
\tilde{h}_{\rm B}^{xyz} c_{\r,\rB}  c_{\r - {\bm a}_1,\rB}
\Bigr], 
\label{eq:Hp}
\end{align}
where $c_{\r,\rA(\rB)}$ is the Majorana fermion operator at the A(B) sublattice site on the $z$ bond at the position $\bm{r}$, and $\tilde{h}_{\rm A(B)}^{\a \b \g} = h_{\rm A(B)}^\a h_{\rm B(A)}^\b h_{\rm A(B)}^\g $; ${\bm a}_1= (1/2, \sqrt{3}/2)$ and ${\bm a}_2 = (-1/2,\sqrt{3}/2)$ are the primitive translational vectors and ${\bm a}_3 \equiv {\bm a}_2 - {\bm a}_1$ (we set the lattice constant as unity) [see Fig.~\ref{fig:honeycomb}(a)]. 
Hereafter, we consider the case with the isotropic ferromagnetic Kitaev couplings, $J^x=J^y=J^z=J > 0$, while the results are the same for the antiferromagnetic case within the above perturbation theory. 
We note that the second-order perturbation with respect to the magnetic field modulates $J^\a$, and the third-order perturbation gives rise to additional three-spin terms to Eq.~\eqref{eq:Hdash} which correspond to interactions between the Majorana fermions. 
We will discuss these additional effects in Sec.~\ref{subsec:Majorana interaction}.

\section{Majorana excitation spectra}
\label{sec:Band}

\begin{figure}
\hspace*{-1mm}
  \includegraphics[width=87mm]{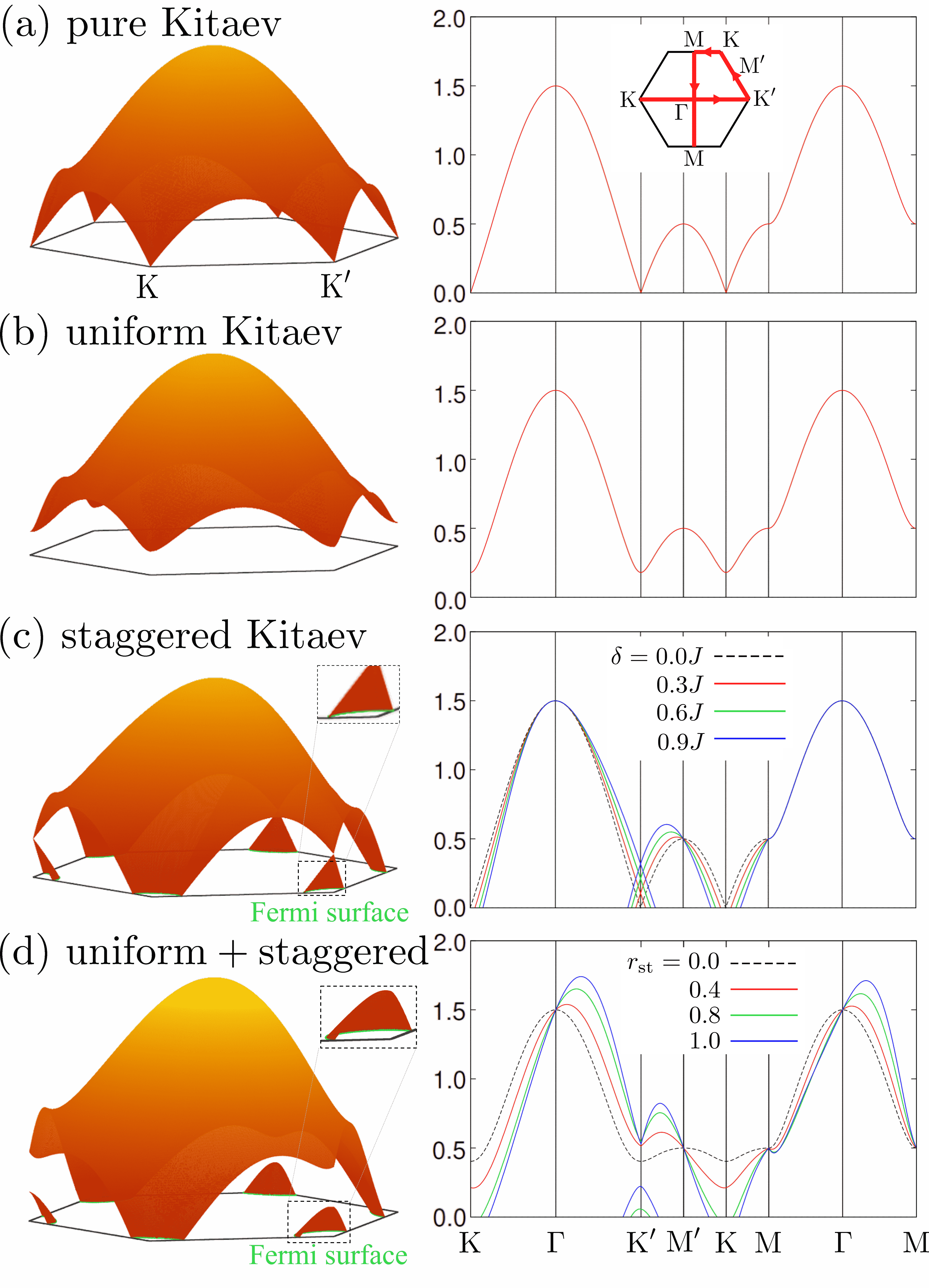}
 \caption{
Majorana band dispersions $E_{\k \pm}/J$ in Eq.~\eqref{eq:eigenenergy}: 
(a) 
in the absence of the magnetic field, 
(b) 
under a uniform magnetic field only $(\delta=0.36J, \ \theta=\varphi=0)$, 
(c) 
under a staggered magnetic field only $(\delta=0.36J, \ \phi_{\rm st}=0)$, and 
(d) 
in the presence of both uniform and staggered fields $(\delta=0.9J, \ \phi_{\rm st}=\pi/6, \ \theta=\pi/3, \ \varphi=0)$. 
The left panels are the bird's-eye views and the right panels are the two-dimensional plots along the symmetric lines 
in the first Brillouin zone depicted in the inset of (a).  
The green lines in the left panels of (c) and (d) denote the Majorana Fermi surfaces. 
The right panels of (c) and (d) include the data while changing $\delta$ and $r_{\rm st}$, respectively. 
}
 \label{fig:band}
\end{figure}

\begin{figure}
  \includegraphics[width=88mm]{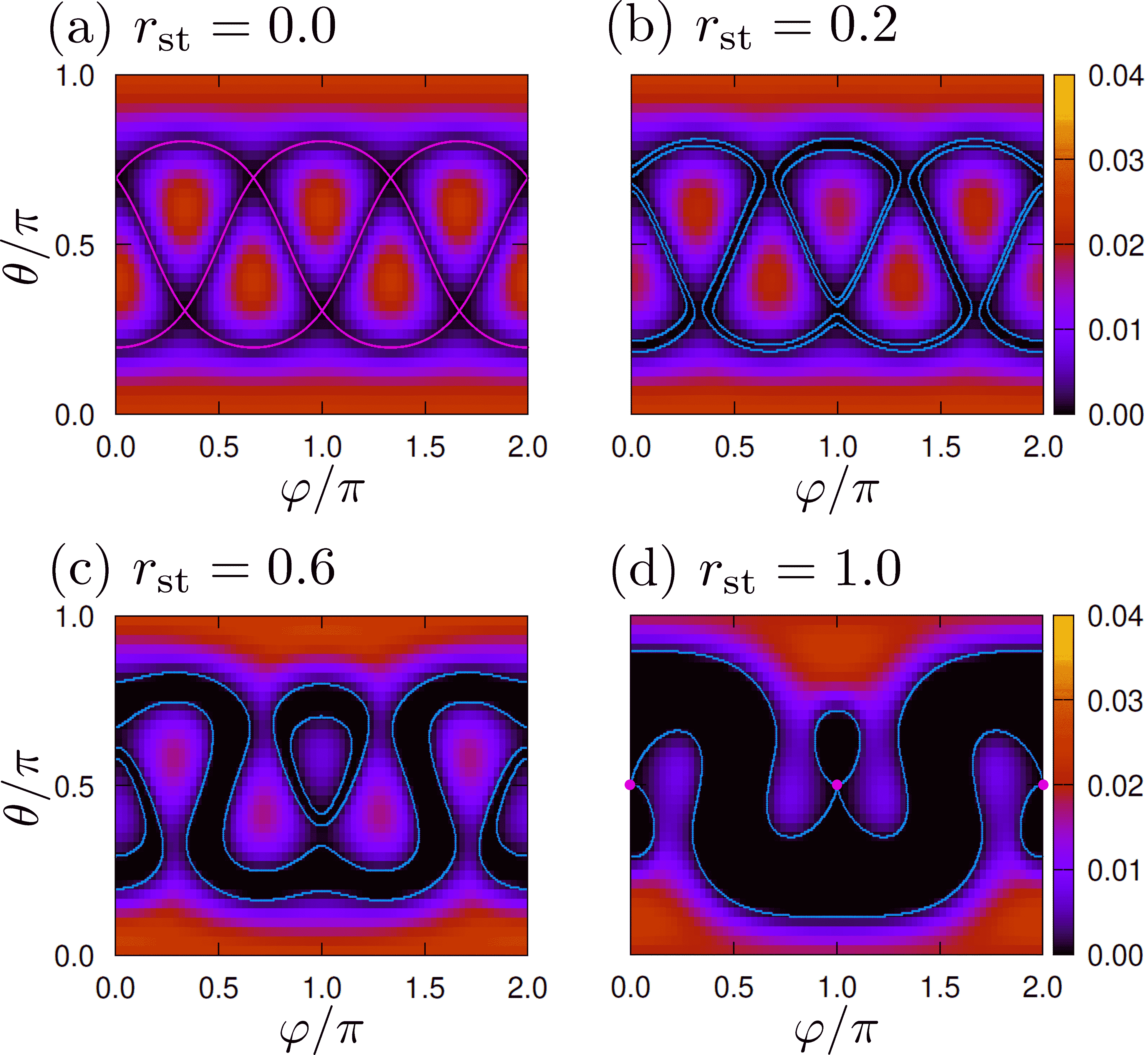}
 \caption{
Contour plots of the Majorana gap $\Delta_{\rm M}/J$ while varying the direction of the uniform component of the magnetic field, ${\bm h}$, with the fixed staggered one ${\bm h}_{\rm st}$ along the $a$ direction $(\phi_{\rm st} = 0)$ for 
(a) $r_{\rm st} = 0$,  
(b) $r_{\rm st} = 0.2$, 
(c) $r_{\rm st} = 0.6$, and 
(d) $r_{\rm st} = 1.0$. 
The cyan lines represent the boundary between gapped and gapless region, while the magenta lines in (a) and points in (d) represent the parameters where the Majorana dispersion has the nodal points at zero energy as in the pure Kitaev model in Fig.~\ref{fig:honeycomb}(a).
We set $\delta = 0.096J$. 
}
 \label{fig:gap}
\end{figure}

\begin{figure}
  \includegraphics[width=85mm]{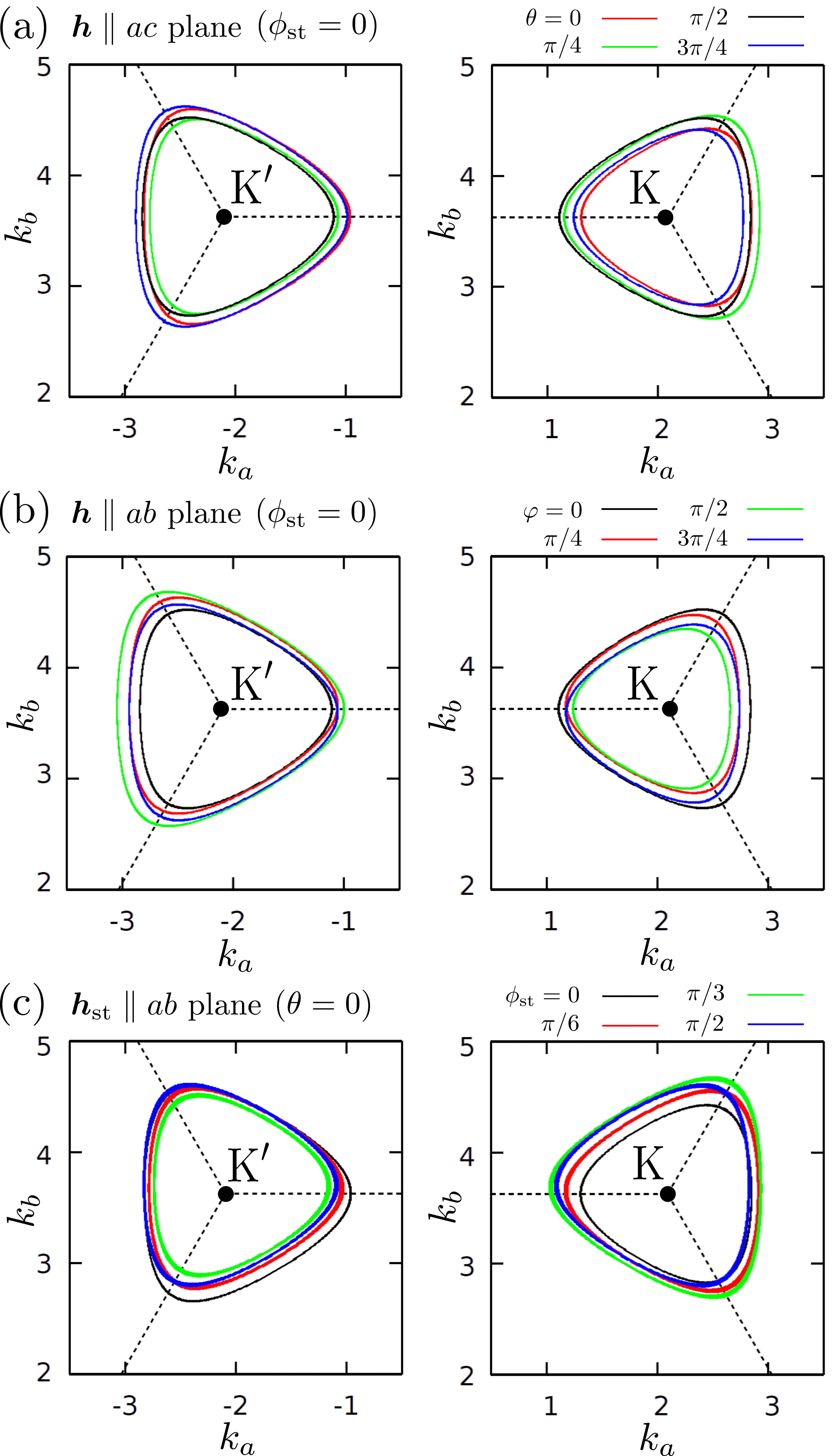}
 \caption{ 
Constant energy cuts at $E_{\k\pm} = 0.35J$ around the K and K$'$ points. 
In (a) and (b), the direction of ${\bm h}$ is varied within the 
(a) $ac$ and  (b) $ab$ planes, while ${\bm h}_{\rm st}$ is fixed along the $a$ direction. 
In (c), the direction of ${\bm h}_{\rm st}$ is varied within the $ab$ plane, while ${\bm h}$ is fixed along the $c$ direction. 
In all the cases, we set $r_{\rm st} = 1$ and $\delta = 0.096J$. 
}
\label{fig:es}
\end{figure}

In this section, we discuss the Majorana excitation spectra in the effective Hamiltonian ${\cal H}_{\rm eff} = {\cal H}_{\rm K} + {\cal H}'$. 
The Fourier transform of Eqs.~\eqref{eq:H_K_Maj} and \eqref{eq:Hp} is summarized into 
\begin{align}
{\cal H}_{\rm eff} = \sum_{\k \in {\rm BZ} ( k_a > 0)}
{\bm c}_\k^\dagger 
\left(
\begin{array}{cc} 
\Delta_{\k,\rA} & if_\k \\
-if_\k^* & \Delta_{\k,\rB} 
\end{array}
\right)
{\bm c}_\k ,
\label{eq:Heff}
\end{align}
where the summation of the momentum ${\bm k}=(k_a,k_b)$ is taken over a half of the first Brillouin zone, and ${\bm c}_\k^\dagger = \left( c_{\k,\rA}^\dg, c_{\k,\rB}^\dg \right)$ and ${\bm c}_\k = {}^t \left( c_{\k,\rA}, c_{\k,\rB} \right)$ are the Fourier components of the Majorana fermion operators defined as
\begin{align} 
c_{\k,\rA(\rB)}^\dg = \frac{1}{\sqrt{2N_{\rm u}}} \sum_\r c_{\r,\rA(\rB)} \expo^{i\k \cdot \r},  
\label{eq:ckdg}
\\
\ \ 
c_{\k,\rA(\rB)} = \frac{1}{\sqrt{2N_{\rm u}}} \sum_\r c_{\r,\rA(\rB)} \expo^{-i\k \cdot \r}, 
\label{eq:ck}
\end{align}
where $N_{\rm u}$ is the number of the unit cell, and
\begin{align}
f_\k &= \frac{J}{2}\left( \expo^{i\k \cdot {\bm a}_1} + \expo^{i\k \cdot {\bm a}_2} + 1 \right) ,
\\
\Delta_{\k,\rA} &=
\frac{\delta}{2} 
\left( \tilde{h}_{\rm A}^{xyz} \sin \k \cdot {\bm a}_1 
      - \tilde{h}_{\rm A}^{zxy} \sin \k \cdot {\bm a}_2 
      + \tilde{h}_{\rm A}^{yzx} \sin \k \cdot {\bm a}_3 
\right), 
\\
\Delta_{\k,{\rm B}} &=
-\frac{\delta}{2} 
\left( \tilde{h}_{\rm B}^{xyz} \sin \k \cdot {\bm a}_1 
      - \tilde{h}_{\rm B}^{zxy} \sin \k \cdot {\bm a}_2 
      + \tilde{h}_{\rm B}^{yzx} \sin \k \cdot {\bm a}_3 
\right).  
\end{align} 
Note that the Majorana operators Eqs.~(\ref{eq:ckdg}) and (\ref{eq:ck}) satisfy $c_{\k,\rA(\rB)}^\dg = c_{-\k,\rA(\rB)}$.  
Then, the eigenenergy is given by the diagonalization of Eq.~\eqref{eq:Heff} as 
\begin{align}
E_{\k \pm} = \frac{1}{2} \left( \Delta_{\k +} \pm \sqrt{ 4|f_\k|^2 + \Delta_{\k -}^2 } \ \right) , 
\label{eq:eigenenergy} 
\end{align} 
with $\Delta_{\k \pm} \equiv \Delta_{\k,\rA} \pm \Delta_{\k,\rB}$. 
Note that $E_{\k \pm}$ is always particle-hole symmetric with respect to zero energy, namely $E_{\k \pm} = -E_{-\k \mp}$, because of the nature of the Majorana fermions, and the only positive energy parts contribute to the physical properties. 

Let us first discuss the overall behavior of the dispersion relation in Eq.~(\ref{eq:eigenenergy}) while changing the uniform and staggered components of the magnetic fields. 
In the absence of the magnetic field $H=0$, $E_{\k \pm}$ has the Dirac-like linear dispersions with the gapless nodal points at the K and K' points, as shown in Fig.~\ref{fig:band}(a)~\cite{Kitaev2006}. 
When introducing a uniform magnetic field, the nodal points are gapped out, as shown in Fig.~\ref{fig:band}(b). 
In this situation, the insulating state becomes topologically nontrivial having a nonzero Chern number~\cite{Kitaev2006}. 
Meanwhile, when we apply a staggered magnetic field only, the nodal points remain gapless, but their energies at the K and K' points are shifted in an opposite energy direction, as shown in Fig.~\ref{fig:band}(c). 
This makes the two valleys at the K and K$'$ points inequivalent and $E_{\k \pm}$ asymmetric in momentum space, 
because of the breaking of sublattice symmetry. 
As a consequence, the staggered magnetic field yields the Majorana Fermi surfaces around the K and K' points, whose sizes are increased with increasing the strength of the staggered magnetic field, as shown in the right panel of Fig.~\ref{fig:band}(c). 
Finally, when we apply both uniform and staggered fields, the gap opening and the asymmetric energy shift occur simultaneously, as shown in Fig.~\ref{fig:band}(d). 
In this case, the system can be insulating or gapless with the Majorana Fermi surfaces depending on the magnitudes of the uniform and staggered components, as demonstrated in the right panel of Fig.~\ref{fig:band}(d). 

Figure~\ref{fig:gap} displays the Majorana gap $\Delta_{\rm M}$ while varying the direction of the uniform component of the magnetic field, ${\bm h}$, with a staggered component ${\bm h}_{\rm st}$ along the $a$ direction. 
When the staggered field is zero or small, the Majorana excitation spectrum is gapped in most cases, as shown in Figs.~\ref{fig:gap}(a) and \ref{fig:gap}(b). 
In the gapped region, the system is the Majorana Chern insulator as in the case with the uniform magnetic field only. 
With an increase of the staggered field, the asymmetric energy shift at the K and K' points becomes larger, 
leading to an increase of the gapless region, as shown in Figs.~\ref{fig:gap}(c) and \ref{fig:gap}(d). 

Figure~\ref{fig:es} shows the asymmetric modulation of $E_{\k \pm}$ by plotting the constant energy cuts at $E_{\k \pm}=0.35J$ around the K and K' points. 
In Fig.~\ref{fig:es}(a), we present the results while rotating ${\bm h}$ within the $ac$ plane with ${\bm h}_{\rm st}$ along the $a$ direction.  
When ${\bm h}$ is parallel to ${\bm h}_{\rm st}$ ($\theta=\pi/2$), the effective magnetic field $\tilde{h}_{A(B)}^{\alpha\beta\gamma}$ in Eq.~\eqref{eq:Hp} becomes zero, and hence, the dispersion is equivalent to that in the absence of the magnetic field and the constant energy cuts are symmetric between the K and K' points. 
The rotation of the direction of ${\bm h}$ modulates the dispersion in an asymmetric way; when the constant energy cut around the K' point is enlarged, that around the K point is shrunk. 
We note that the asymmetry is always present in the $k_a$ direction except for $\theta = \pi/2$, while the dispersion becomes symmetric in the $k_b$ direction for $\theta = 0$, $\pi/4$, and $3\pi/4$.  
On the other hand, when ${\bm h}$ is rotated within the $ab$ plane, the dispersion becomes asymmetric in both $k_a$ and $k_b$ directions, except for the symmetric case with $\varphi=0$, as shown in Fig.~\ref{fig:es}(b). 
Finally, Fig.~\ref{fig:es}(c) shows the results while rotating ${\bm h}$ within the $ab$ plane with ${\bm h}_{\rm st}$ along the $c$ direction $(\theta=0)$, which demonstrates that the asymmetry is changed by the rotation of ${\bm h}_{\rm st}$.

\section{Thermal transport}
\label{sec:thermal}

In this section, we calculate the longitudinal thermal conductivity in the presence of both uniform and staggered magnetic fields, by using the Boltzmann transport theory. 
We apply a weak thermal gradient to the $a$ or $b$ direction. 
Then, the thermal current can be expanded in terms of the thermal gradient as
\begin{align}
j_{Q,\xi} = \kappa_\xi^{(1)} \left( -\frac{\partial T}{\partial \xi} \right) + \kappa_\xi^{(2)} \left( \frac{\partial T}{\partial \xi} \right)^2 + \cdots
\label{eq:j_Q}
\end{align}
where $\xi=a$ or $b$; $\kappa_\xi^{(1)}$ and $\kappa_\xi^{(2)}$ are the linear and nonlinear thermal conductivities given by 
\begin{align}
\label{eq:ltc}
\kappa_\xi^{(1)} &= \frac{2\tau}{\Omega} \sum_n  \sum_{\k \in {\rm BZ} (E_{\k n}>0) } 
\frac{ \partial f (E_{\k n})}{ \partial T}  E_{\k n} v_{\k n \xi}^2, \\
\label{eq:nltc}
\kappa_\xi^{(2)} &= \frac{2 \tau^2}{\Omega} \sum_n \sum_{\k \in {\rm BZ} (E_{\k n}>0) } 
\frac{ \partial^2 f(E_{\k n}) }{ \partial T^2}  E_{\k n} v_{\k n \xi}^3, 
\end{align}
respectively, where the summation is taken over the momentum ${\bm k}$ in the first Brillouin zone for which $E_{{\bm k}n}>0$, $\Omega$ is the system size, $n = \pm$ is the band index, $f(E_{\k n}) = (\expo^{E_{\k n}/T} + 1)^{-1}$ is the Fermi distribution function, $v_{\k n \xi} \equiv \partial E_{\k n}/\partial k_{\xi} $ is the group velocity, and $\tau$ is the relaxation time of Majorana fermion introduced phenomenologically. 
We set the Boltzmann constant $k_{\rm B}$ unity and take $J \tau=1$ in this section. 

The thermal current is generated by thermally excited Majorana fermions. To discuss the field dependence of the linear response $\kappa_\xi^{(1)}$, we introduce the ``effective Majorana density" as
\begin{align}
D_{\rm M} = \int_0^\infty \sum_n \sum_{\k \in {\rm BZ} (E_{\k n}>0)} \delta (\varepsilon - E_{\k n}) \left( - \frac{df(\varepsilon)}{d\varepsilon} \right) d\varepsilon ,
\end{align} 
in addition to the Majorana gap $\Delta_{\rm M}$; see Sec.~\ref{subsec:linear}.
Meanwhile, $\kappa_\xi^{(2)}$ describes the second-order nonlinear response arising from the asymmetry of the Majorana band dispersions. 
To measure the asymmetry, we define the \lq\lq band asymmetry parameter'' as 
\begin{align}
\L_\xi 
= -\frac{1}{\Omega}  
\sum_{\k \in {\rm BZ} ( k_\xi > 0 )} \sum_n \frac{\partial f_{\k n}}{ \partial E_{\k n} } 
( v_{\k n \xi} + v_{-\k n \xi} ) ,
\label{eq:asp}
\end{align} 
where the sum of momentum is taken in half of the first Brillouin zone with $k_\xi>0$ to extract the asymmetry. 
We use this quantity to discuss the field dependences of $\kappa_\xi^{(2)}$ in Sec.~\ref{subsec:nonlinear}. 
We perform the following calculations for the system with periodic boundary conditions by taking $864 \times 864$ $k$ points in the first Brillouin zone. 

\subsection{Linear thermal conductivity}
\label{subsec:linear}

\begin{figure}[t]
\hspace*{-1.8mm}
  \includegraphics[width=84mm]{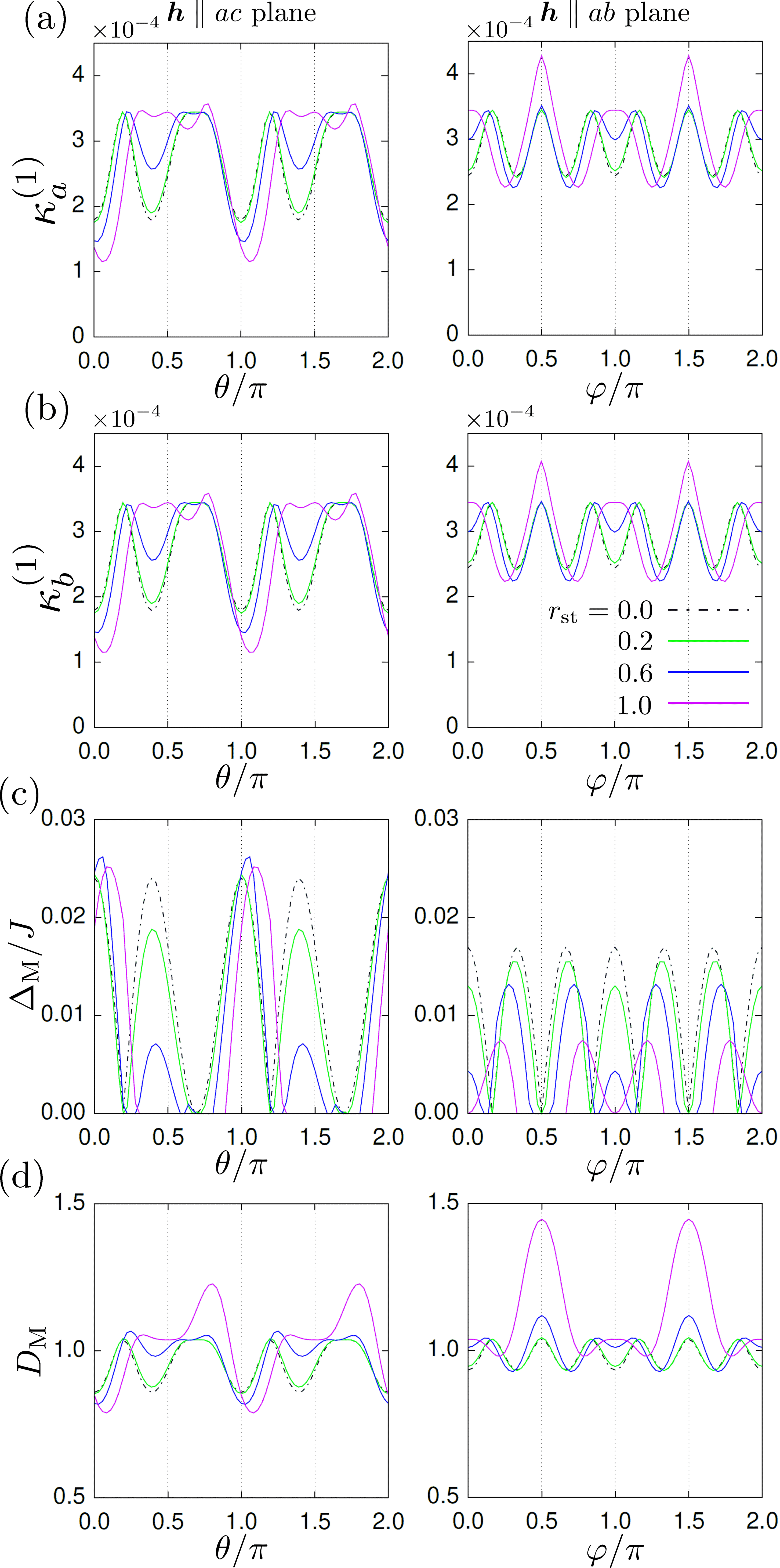}
\vspace*{-2mm}
 \caption{ 
The linear thermal conductivities 
(a) $\kappa_{a}^{(1)}$ and (b) $\kappa_{b}^{(1)}$,  
(c) the Majorana gap $\Delta_{\rm M}$, 
and (d) the integrated density of states $D_{\rm M}$ 
while varying the direction of the uniform magnetic field in the $ac$ plane (left panels) and the $ab$ plane (right panels) under the staggered magnetic field along the $a$ direction $(\phi_{\rm st}=0)$.  
The dot-dashed black lines represent the data for the uniform magnetic field only ($r_{\rm st}=0$), while the green, blue, and magenta lines are for the cases with the staggered magnetic field at $r_{\rm st} = 0.2$, $0.6$, and $1.0$, respectively. 
We set $\delta = 0.096J$ and temperature $T=0.01J$. 
}
 \label{fig:thcondl}
\end{figure}

\begin{figure*}[t]
  \includegraphics[width=180mm]{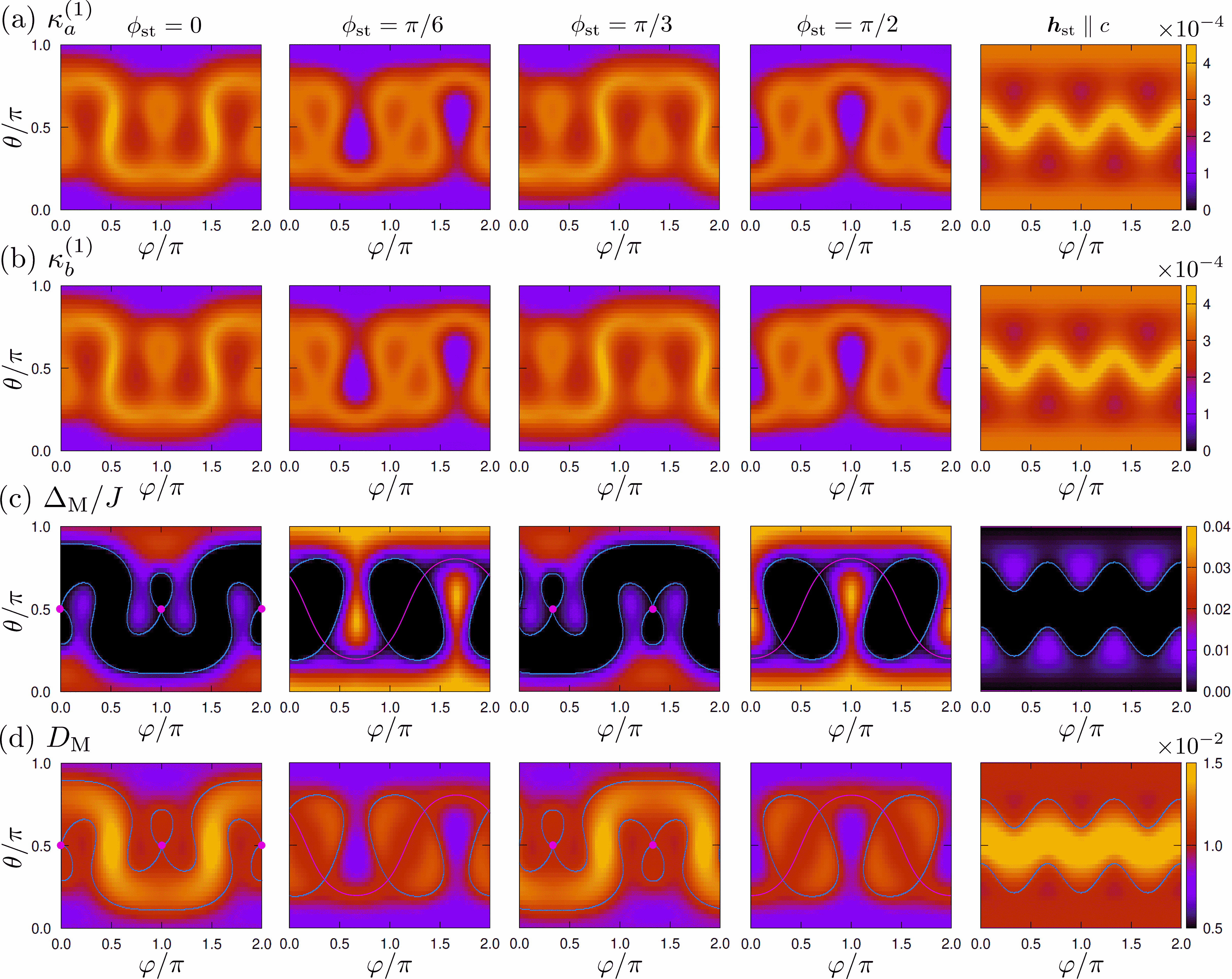}
 \caption{ 
 Contour plots of 
the linear thermal conductivities (a) $\kappa_{a}^{(1)}$ and (b) $\kappa_{b}^{(1)}$, (c) the Majorana gap $\Delta_{\rm M}$, and (d) the effective Majorana density $D_{\rm M}$ while varying the direction of the uniform magnetic field under the 
staggered magnetic field with $\phi_{\rm st}=0$, $\pi/6$, $\pi/3$, $\pi/2$, and ${\bm h}_{\rm st} \parallel c$ from left to right. 
In (c) and  (d), the cyan lines and the magenta lines and points are drawn with the same notations as Fig.~\ref{fig:gap}.  
We set $\delta = 0.096J$, $r_{\rm st} = 1$, and $T=0.01J$. 
}
 \label{fig:thcond_colorl}
\end{figure*}

First, we show the results for the linear thermal conductivities $\kappa_{a}^{(1)}$ and $\kappa_{b}^{(1)}$ in Figs.~\ref{fig:thcondl}(a) and \ref{fig:thcondl}(b), respectively, while rotating the uniform magnetic field in the $ac$ and $ab$ planes. 
We find that they change drastically with the field angle and the angle dependence becomes more conspicuous for a larger staggered magnetic field (namely, larger $r_{\rm st}$), while $\kappa_{a}^{(1)}$ and $\kappa_{b}^{(1)}$ are almost identical. 
To understand these behaviors, we plot the Majorana gap $\Delta_{\rm M}$ in Fig.~\ref{fig:thcondl}(c). 
When the staggered magnetic field is absent ($r_{\rm st}=0.0$) or relatively weak ($r_{\rm st}=0.2$), the Majorana spectrum changes with the field angle, while it is gapped in most regions of $\theta$ and $\varphi$. 
Meanwhile, when the staggered magnetic field becomes larger and comparable to the uniform one ($r_{\rm st} = 0.6$ and $1.0$), the spectrum becomes gapless and the Majorana Fermi surfaces appear in wider regions, as shown in Fig.~\ref{fig:thcondl}(c). 
In the gapped region, the angle dependence of $\kappa_\xi^{(1)}$ appears to inversely correlate with $\Delta_{\rm M}$, as expected for the thermal transport in the gapped system. 
On the other hand, as shown in Fig.~\ref{fig:thcondl}(d), in the gapless region, we find that $\kappa_\xi^{(1)}$ roughly correlates with the effective Majorana density $D_{\rm M}$. 
For example, for $r_{\rm st} = 1$, we observe that $D_{\rm M}$ is almost angle independent in the region where $\kappa_\xi^{(1)}$ does not change so much in the case of ${\bm h} \parallel ac$ plane, while $D_{\rm M}$ increases around $\varphi/\pi = 0.5$ and $1.5$ where $\kappa_\xi^{(1)}$ is enhanced in the case of ${\bm h} \parallel ab$ plane. 

In Fig.~\ref{fig:thcond_colorl}, we display $\kappa_\xi^{(1)}$ in comparison with $\Delta_{\rm M}$ in a more comprehensive manner while changing the direction of the staggered magnetic field $\phi_{\rm st}$. 
We take the amplitudes of the uniform and staggered magnetic fields to be the same, that is, $r_{\rm st} = 1$. 
Again, we observe inverse correlation between $\kappa_\xi^{(1)}$ and $\Delta_{\rm M}$. 
In addition, we find that $\kappa_\xi^{(1)}$ at $\phi_{\rm st} = 0$ and $\phi_{\rm st} = \pi/6$ behave similarly to those at $\phi_{\rm st} = \pi/3$ and $\phi_{\rm st}=\pi/2$, respectively, with a shift of $\varphi$. 
This is because the Majorana gap satisfies the relation
\begin{align}
\Delta_{\rm M} (\phi_{\rm st},\varphi) = \Delta_{\rm M} \left(\phi_{\rm st}+\frac{m\pi}{3},\varphi-\frac{2m\pi}{3}\right) ,
\end{align}
where $m$ is an integer. 
Note, however, that the values of $\kappa_\xi^{(1)} (\phi_{\rm st},\varphi)$ and $\kappa_\xi^{(1)} (\phi_{\rm st} + m\pi/3,\varphi-2m\pi/3)$ are slightly different because the Majorana spectra are different. 
As plotted in the rightmost panels in Fig.~\ref{fig:thcond_colorl}, the results for the out-of-plane staggered magnetic field show threefold rotational symmetry for $\varphi$, in contrast to those for the in-plane ones.

\subsection{Nonlinear thermal conductivity}
\label{subsec:nonlinear}

\begin{figure}
  \includegraphics[width=84mm]{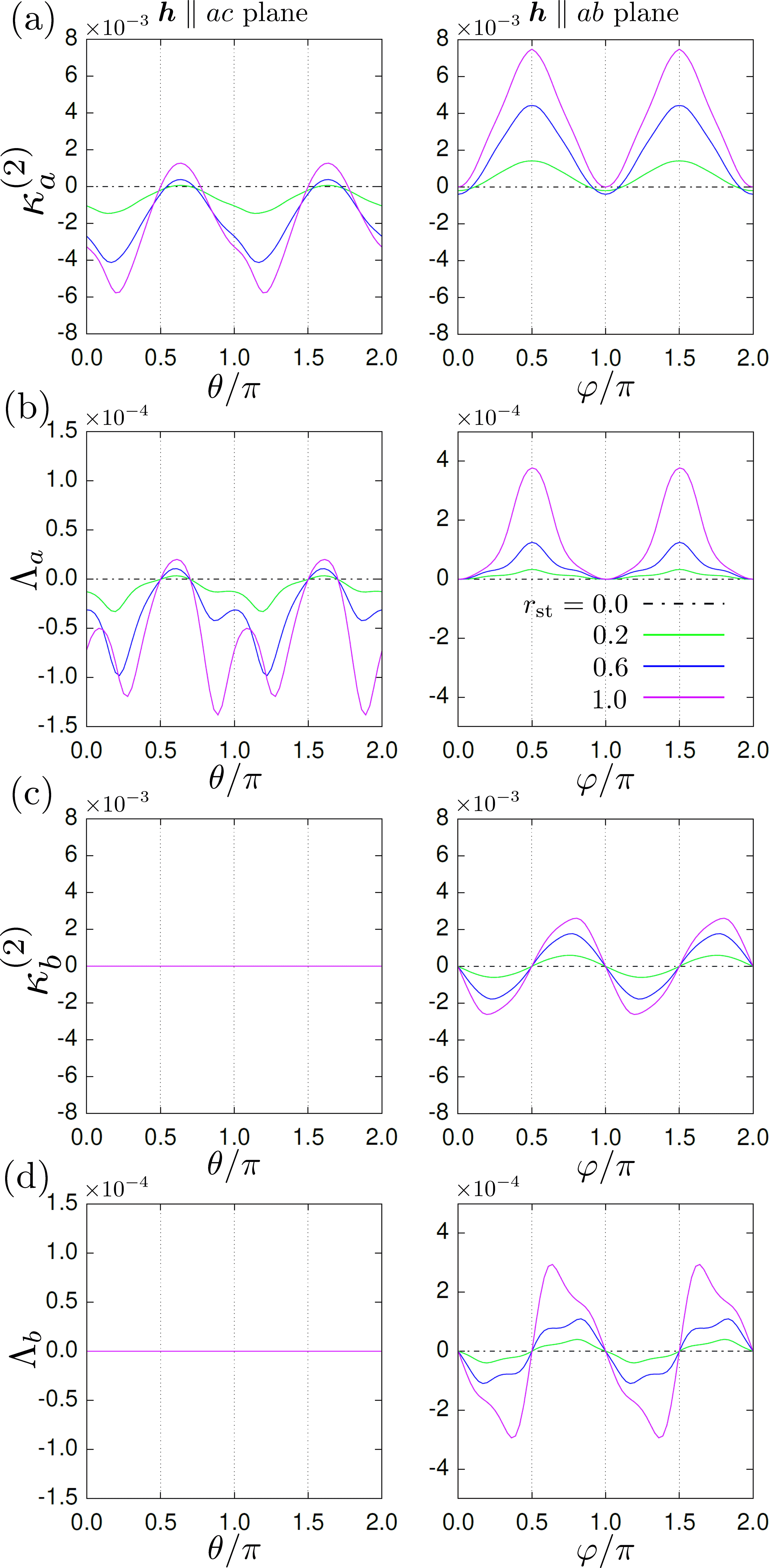}
 \caption{ 
The nonlinear thermal conductivities (a) $\kappa_{a}^{(2)}$ and (c) $\kappa_{b}^{(2)}$, and the band asymmetry parameters (b) $\L_a$ and (d) $\L_b$ for the same parameter settings as those in Fig.~\ref{fig:thcondl}. 
}
 \label{fig:thcondnl}
\end{figure}

\begin{figure*}[t]
  \includegraphics[width=180mm]{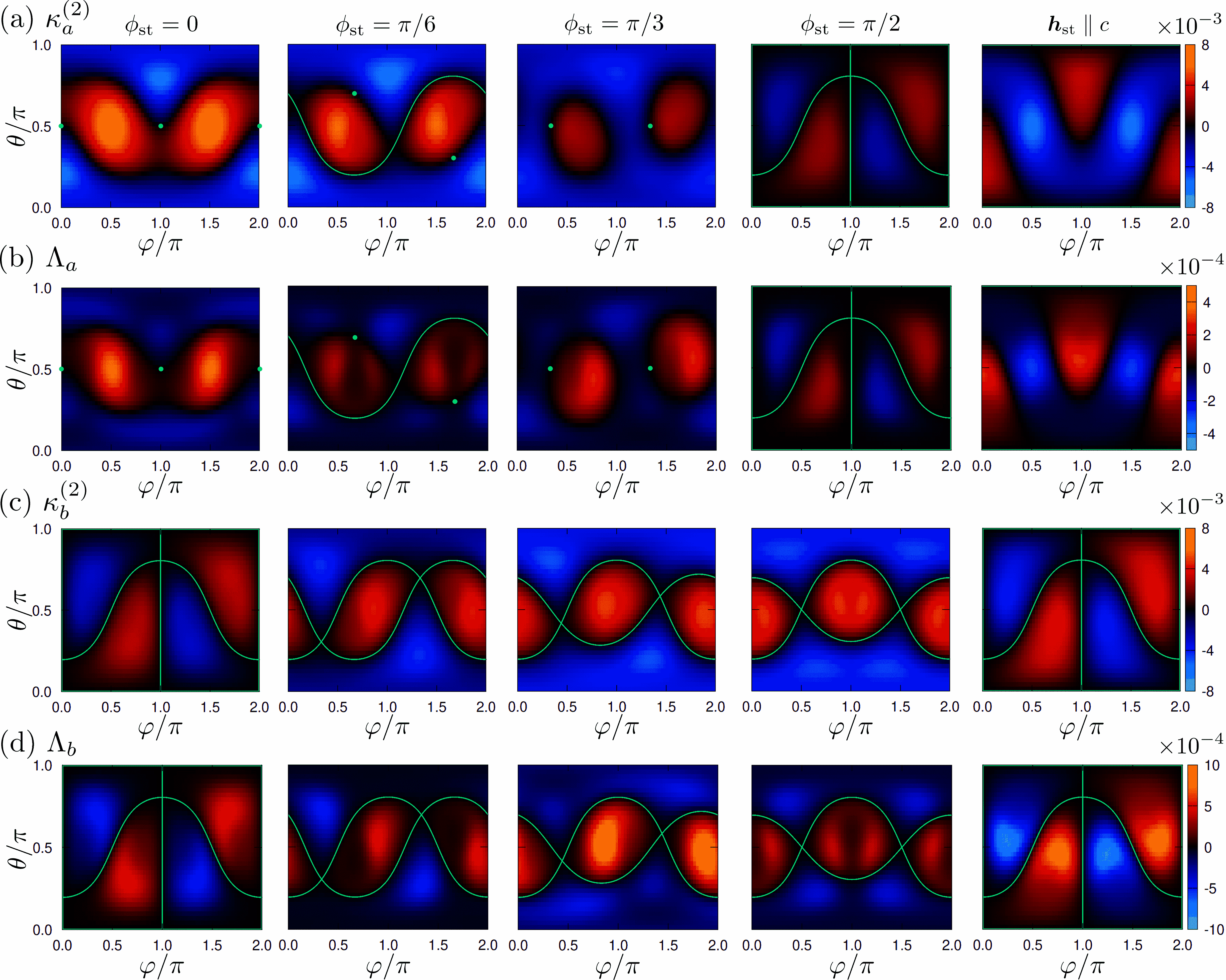}
 \caption{ 
Contour plots of 
the nonlinear thermal conductivities (a) $\kappa_{a}^{(2)}$ and (c) $\kappa_{b}^{(2)}$, and the band asymmetry parameters (b) $\L_a$ and (d) $\L_b$ for the same parameter settings as those in Fig.~\ref{fig:thcond_colorl}.  
The green lines and points represent the parameters where the Majorana excitation spectrum becomes symmetric for the given magnetic fields; see the text for details. 
}
 \label{fig:thcond_colornl}
\end{figure*}

Next, we show the results for the nonlinear thermal conductivities $\kappa_{a}^{(2)}$ and $\kappa_{b}^{(2)}$ in Figs.~\ref{fig:thcondnl}(a) and \ref{fig:thcondnl}(c), respectively, while rotating the uniform magnetic field in the $ac$ and $ab$ planes. 
In contrast to $\kappa_\xi^{(1)}$, $\kappa_\xi^{(2)}$ vanishes when the staggered magnetic field is absent ($r_{\rm st}=0.0$), as the Majorana band dispersions are symmetric in momentum space. 
Upon introducing the staggered magnetic field, the Majorana dispersions become asymmetric, and $\kappa_\xi^{(2)}$ becomes nonzero and show the field angle dependence. 
Interestingly, we find that $\kappa_\xi^{(2)}$ exhibits distinct angle dependence between $\kappa_{a}^{(2)}$ and $\kappa_{b}^{(2)}$; in particular, $\kappa_{b}^{(2)}$ is always zero when the uniform field is rotated in the $ac$ plane, as plotted in the left panel of Fig.~\ref{fig:thcondnl}(c). 
This is in sharp contrast to the behaviors of $\kappa_{a}^{(1)}$ in Fig.~\ref{fig:thcondl}. 
We compare these behaviors with the band asymmetric parameter $\L_\xi$ in Eq.~(\ref{eq:asp}) plotted in Figs.~\ref{fig:thcondnl}(b) and \ref{fig:thcondnl}(d). 
As expected, we find clear correlation between $\kappa_\xi^{(2)}$ and $\L_\xi$ in their dependences on the field angle and $r_{\rm st}$. 
In particular, $\L_b$ vanishes for the uniform field in the $ac$ plane because of the mirror symmetry, which results in zero $\kappa_{b}^{(2)}$. 
Our result suggests that the Majorana band asymmetry can be detected by the measurement of nonreciprocal 
thermal transport, while higher experimental accuracy is needed as discussed later. 

Figure~\ref{fig:thcond_colornl} displays $\kappa_\xi^{(2)}$ and $\L_\xi$ in a similar manner to Fig.~\ref{fig:thcond_colorl}. 
We again observe the overall correlation between $\kappa_\xi^{(2)}$ and $\L_\xi$. 
Moreover, from the comparison with Fig.~\ref{fig:thcond_colorl}(c), we find that $\kappa_\xi^{(2)}$ is further enhanced in the gapless states with the Majorana Fermi surfaces. 
We note that $\kappa_a^{(2)}$ and $\L_a$ look differently at $\phi_{\rm st}=\pi/6$: $\L_a$ has the dips around $\phi=0.7\pi$ and $1.7\pi$, but $\kappa_a^{(2)}$ does not.  
Similar discrepancies can be seen also for $\kappa_b^{(2)}$ and $\Lambda_b$ at $\phi_{\rm st}=\pi/2$. 
We speculate that these behaviors reflect different temperature effects on $\L_a$ and $\kappa_a^{(2)}$. 

We also note that $\kappa_\xi^{(2)}$ as well as $\L_\xi$ vanishes on several lines and points denoted by green in Fig.~\ref{fig:thcond_colornl}. 
On these lines and points, the Majorana excitation spectrum in Eq.~\eqref{eq:eigenenergy} becomes symmetric in momentum space, and hence, the band asymmetry vanishes. 
Specifically, Eq.~\eqref{eq:eigenenergy} becomes symmetric with respect to the $a$ direction when $\tilde{h}_{\rm A}^{xyz} - \tilde{h}_{\rm B}^{xyz} = - ( \tilde{h}_{\rm A}^{zxy} - \tilde{h}_{\rm B}^{zxy} )$ and $\tilde{h}_{\rm A}^{yzx} - \tilde{h}_{\rm B}^{yzx} = 0$, while it is symmetric with respect to the $b$ direction when $\tilde{h}_{\rm A}^{xyz} - \tilde{h}_{\rm B}^{xyz} = \tilde{h}_{\rm A}^{zxy} - \tilde{h}_{\rm B}^{zxy}$.

\section{Discussion}
\label{sec:disc}

\subsection{Quantitative estimate}
\label{subsec:order}

Let us estimate quantitatively the linear and nonlinear thermal currents, $j_{Q,\xi}^{(1)}$ and $j_{Q,\xi}^{(2)}$, which are given by the first and second terms in Eq.~\eqref{eq:j_Q}, respectively. 
Bearing the candidate material $\a$-$\rm RuCl_3$ in mind, we set the in-plane and out-of-plane lattice constants as $a_0 = 5.98$~$\ong$ and $c_0 = 5.72$~$\ong$, respectively~\cite{Johnson2015}.  
In the following calculations, we use $c_0$ to take the $k_c$ integral from 0 to $2\pi/c_0$ assuming no dispersion in the $k_c$ direction. 
We also assume the Kitaev interaction $J/k_{\rm B} = 80$~K~\cite{Sandilands2015,NKKMM}. 
We here set the temperature gradient $\partial T/\partial a = 160~{\rm K/m}$, referring the experimental setup~\cite{Kasahara, Hirobe}.  
We also take the amplitude of the uniform magnetic field at $H/g\mu_{\rm B} \sim 7~{\rm T}$, where $g$ is the electron $g$-factor and $\mu_{\rm B}$ is the Bohr magneton, which roughly corresponds to $\delta \sim 0.1J$ when assuming $\Delta_{\rm f}/k_{\rm B} \sim 10~{\rm  K} $~\cite{Kasahara}. 
\begin{figure}[t]
  \includegraphics[width=85mm]{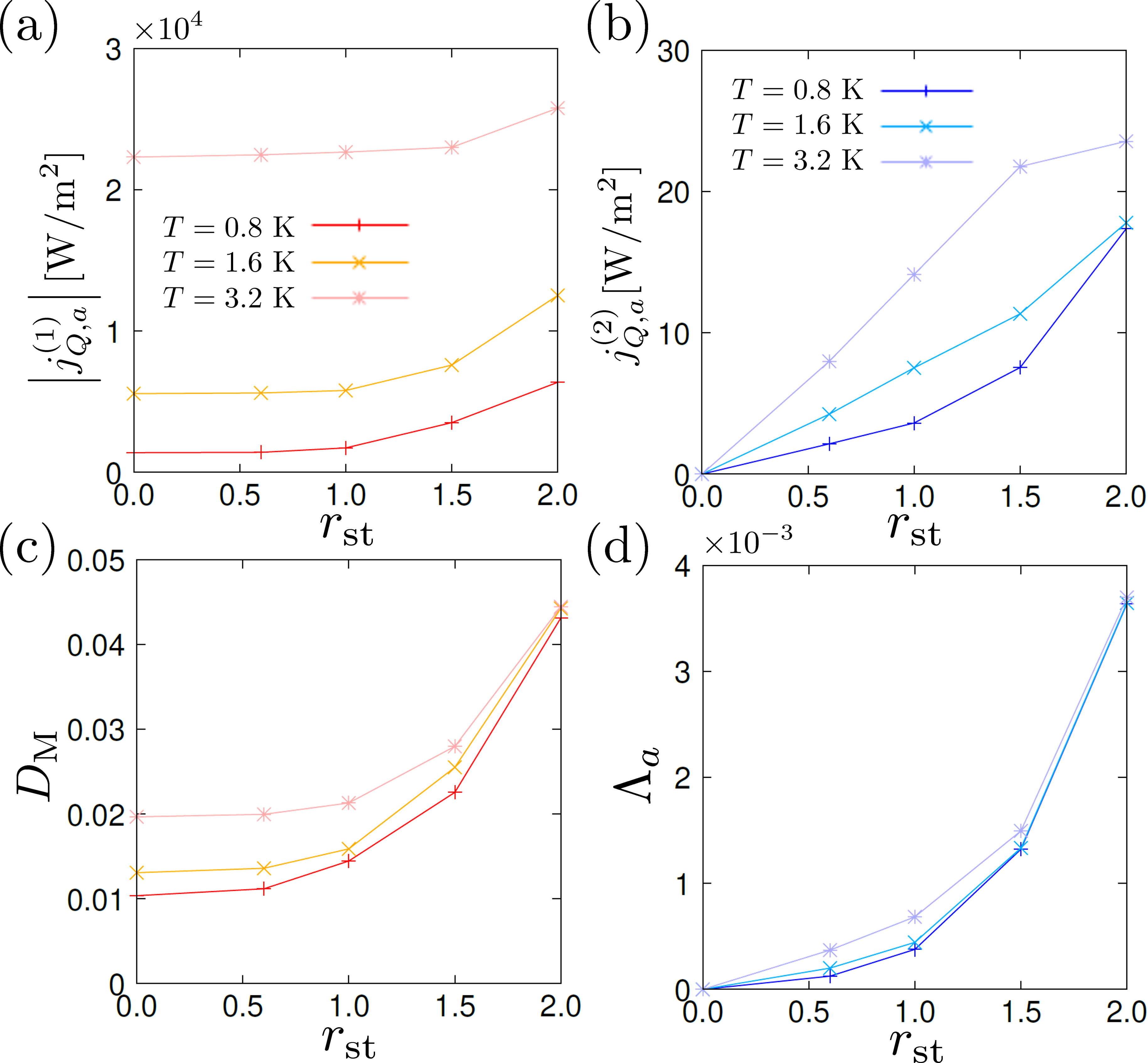}
 \caption{ 
Quantitative estimates of the (a) linear and (b) nonlinear components of thermal currents, under the magnetic field configuration $(\theta,\varphi,{\phi}_{\rm st}) = (\pi/2, \pi/2, 0)$ which means ${\bm h} \parallel b$ and ${\bm h}_{\rm st} \parallel a$, with the uniform magnetic field $H/g\mu_{\rm B}= 6.8$~T. 
For comparison, the effective Majorana density and the band asymmetry parameter are shown in (c) and (d), respectively.
We assume the parameters in $\a$-$\rm RuCl_3$; see the text for details. 
The lifetime is assumed as $J\tau / \hbar = 10^5$. 
}
 \label{fig:qual}
\end{figure}

Figure~\ref{fig:qual}(a) shows the $r_{\rm st}$ dependence of the magnitude of the linear component of the thermal current, $| j_{Q,a}^{(1)}|$, at three different temperatures $T=0.8$, $1.6$, and $3.2$~K in the uniform and staggered magnetic fields applied along the $b$ and $a$ directions, respectively, which leads to the gapless spectrum with the Fermi surfaces around the K and K' points and the nodal points at nonzero energy at the K points, similar to Fig.~\ref{fig:band}(c). 
Note that the sign of $j_{Q,a}^{(1)}$ depends on the direction of the temperature gradient; see Eq.~\eqref{eq:j_Q}. 
We here assume an extremely clean case by taking $J\tau / \hbar = 10^5$. 
We find that $| j_{Q,a}^{(1)}|$ does not depend strongly on $r_{\rm st}$ for $0 < r_{\rm st} \lesssim 1$, whereas rapidly increases for $r_{\rm st} > 1$. 
This behavior is well correlated with the change of the effective Majorana density $D_{\rm M}$ shown in Fig.~\ref{fig:qual}(c). 
We also find that $| j_{Q,a}^{(1)}|$ at $r_{\rm st} \sim 2$ increases almost linearly with $T$, while it increases faster at lower $r_{\rm st}$. 
In contrast, $D_{\rm M}$ depends on temperature for smaller $r_{\rm st}$, while it is almost temperature independent at $r_{\rm st} \sim 2$. 
These behaviors come from the energy dependence of the Majorana density of states (DOS) as follows. 
The Majorana DOS changes its functional form below and above the energy of the nodal points at the K points; 
the DOS is almost constant below the nodal points, while it increases almost linearly above the nodal points. 
When $r_{\rm st}$ is small, the nodal points locate at a low energy and the linearly increasing DOS above the nodal points dominates the thermal transport; hence, $| j_{Q,a}^{(1)}|$ increases faster than $T$-linear. 
On the other hand, when $r_{\rm st}$ becomes large, the nodal points shift to a higher energy and the constant DOS dominates the thermal transport, leading to the $T$-linear dependence of $| j_{Q,a}^{(1)}|$. 
When we take $r_{\rm st} \simeq 1.5$, the estimate of $| j_{Q,i}^{(1)} |$ is in the order of $10^4$~W/m$^2$, namely, $\kappa_{a}^{(1)} \sim 10^2$~W/(K$\cdot$m). Even if we compromise the relaxation time down to $J\tau / \hbar = 10^3$, $\kappa_a^{(1)}$ is estimated as $\sim 1$~W/m$^2$, 
which is comparable to the values in the previous experiments~\cite{Kasahara,Hirobe}. 
Hence, we expect that the characteristic field angle dependence of the linear thermal conductivity in Fig.~\ref{fig:thcond_colorl} is observable. 
In reality, it may compete with the contribution from phonons~\cite{Hirobe,Leahy,Yu,Hentrich,Kasahara,Kasahara2,Hentrich2}, but it would be able to identify the Majorana contribution by the field angle dependence. 

Next, let us discuss the nonlinear component of the thermal current $j_{Q,a}^{(2)}$ for the same field configuration. 
In this situation, the asymmetry of the Majorana dispersion is present only in the $a$ direction.  
As shown in Fig.~\ref{fig:qual}(b), we find that $j_{Q,a}^{(2)}$ increases with the increase of the asymmetry parameter $\Lambda_a$ shown in Fig.~\ref{fig:qual}(d).
We also note that $j_{Q,a}^{(2)}$  depends on $T$ in the small $r_{\rm st}$ region, whereas the $T$ dependence becomes weaker for larger $r_{\rm st}$, because of the energy dependence of the Majorana DOS discussed above. 
The estimate of $j_{Q,a}^{(2)}$ at $r_{\rm st} \simeq 1.5$ and $T=3.2~{\rm K}$ is about $22~{\rm W/m^2}$.  
This value is about 0.4~\% of $| j_{Q,a}^{(1)} |$, which would be hard to observe in experiments at present, and future improvement is desired. 

\subsection{Other additional effects from magnetic fields}
\label{subsec:Majorana interaction}

In the derivation of the effective Hamiltonian in Sec.~\ref{sec:Model}, we omit the contribution from the second-order perturbation in terms of the magnetic fields, 
\begin{align}
{\cal H}_2 &= - \delta_2 \sum_{\a=x,y,z} \sum_{\langle i,j  \rangle_\a} h_i^\a h_j^\a S_i^\a S_j^\a 
\nonumber \\
&= - \frac{i\delta_2}{4} \sum_{\r} \bigl[ \tilde{h}_2^x c_{\r, \rA} c_{\r + {\bm a}_1,\rB} 
+ \tilde{h}_2^y c_{\r,\rA} c_{\r + {\bm a}_2,\rB} + \tilde{h}_2^z c_{\r,\rA} c_{\r,\rB} \bigr] , 
\label{eq:H2}
\end{align}
where $\tilde{h}_2^\a = h_{\rm A}^\a h_{\rm B}^\a$ and $\delta_2$ is assumed to be a constant.  
Equation~\eqref{eq:H2} indicates that the second-order contribution modifies the Kitaev couplings $J^\a$ in an anisotropic way, which would modulate both linear and nonlinear thermal responses as discussed below. 

We also omit the other type of three-spin terms appearing in the third-order perturbation, which is given by~\cite{Kitaev2006} 
\begin{align} 
{\cal H}_4 &= - \delta_4 \sum_{[ \ell , m , n ]} h_\ell^x h_m^y h_n^z S_\ell^x S_m^y S_n^z 
\nonumber \\
&=- \frac{\delta_4}{8} \sum_\r \bigl[ \tilde{h}_{\rm A} c_{\r,\rB} c_{\r,\rA}  c_{\r - {\bm a}_1,\rA}  c_{\r - {\bm a}_2,\rA} 
+ \tilde{h}_{\rm B} c_{\r,\rA} c_{\r,\rB}  c_{\r + {\bm a}_2,\rB}  c_{\r + {\bm a}_1,\rB} \bigr],  
\label{eq:H4}
\end{align} 
where the summation in the first line is taken for three sites $[\ell,m,n]$ on the same sublattice as shown in Fig.~\ref{fig:honeycomb}(a), and $\tilde{h}_{\rm A(B)} = h_{\rm A(B)}^x h_{\rm A(B)}^y h_{\rm A(B)}^z $; 
$\delta_4$ is assumed to be a constant. 
Equation~\eqref{eq:H4} describes interactions between the Majorana fermions, which hamper the exact solvability of the problem. 
At the level of a mean-field approximation, Eq.~\eqref{eq:H4} leads to modifications of not only $J^\a$ but also $\tilde{h}_{\rA(\rB)}^{\a\b\g}$. 
Another possible effect of the Majorana interactions is an instability of the Fermi surfaces, but this issue is not yet fully understood~\cite{CMR}. 

Summarizing Eq.~\eqref{eq:H2} and the mean-field contributions from Eq.~\eqref{eq:H4}, we end up with the one-body Hamiltonian in the form of Eqs.~\eqref{eq:H_K_Maj} and \eqref{eq:Hp} with normalized  $J^\alpha$ and $\tilde{h}_{\rm A(B)}^{\alpha\beta\gamma}$ as 
\begin{align}
\tilde{J}^\a &= 
J^\a 
+ \delta_2 \tilde{h}_2^\a 
- \delta_4 \left( \tilde{h}_{\rm A} \Delta_{\rm AA} + \tilde{h}_{\rm B} \Delta_{\rm BB} \right) ,
\label{eq:Jmod}
\\
\tilde{\tilde{h}}_{\rm A(B)}^{\a \b \g} 
&=
\tilde{h}_{\rm A(B)}^{\a \b \g}
 - \frac{\delta_4}{\delta} \tilde{h}_{\rm A(B)} \Delta_{\rm AB},
\label{eq:kapab} 
\end{align} 
where $\a,\b,\g = x, y, z$; 
here, we assume 
spatially uniform mean fields 
$\Delta_{\rm AA} \equiv i \langle a_{\r,\rA} a_{\r',\rA} \rangle$, 
$\Delta_{\rm BB}\equiv i \langle b_{\r,\rB} b_{\r',\rB} \rangle$, and 
$\Delta_{\rm AB} \equiv i \langle a_{\r,\rA} b_{\r',\rB} \rangle$ 
to decouple the two-body terms in Eq.~\eqref{eq:H4}. 
While $J^\alpha$ are perturbed weakly as $\delta_2, \delta_4 \ll J^\alpha$, $\tilde{h}_{\rm A(B)}^{\alpha\beta\gamma}$ are modified considerably since the first and second terms in Eq.~(\ref{eq:kapab}) can be the same order. 
We note that $\Delta_{\rm AB}$ is estimated as $\sim -0.52$ in the absence of magnetic field~\cite{BMS,NUM,LF}. 
Thus, this argument suggests that we are able to understand the whole magnetic field dependence if we could fit the experimental data by the appropriate tuning of $\delta$, $\delta_2$, and $\delta_4$, though the detailed analysis is left for future study. 

\subsection{Contributions from magnons} 

One way to apply a staggered magnetic field to the Kitaev spin liquid is to make a heterostructure between a Kitaev candidate materials and a honeycomb antiferromagnet, where an internal field from the antiferromagnet acts as a  staggered field. 
Such a situation would be realized in a heterostructure composed of $\alpha$-RuCl$_3$ and a honeycomb antiferromagnet. 
Note that van der Waals heterostructures of atomically thin $\alpha$-${\rm RuCl_3}$ and graphene have been fabricated~\cite{Mashhadi,Zhou,Rizzo,Leeb}, which would be extended to the current situation. 
Besides, a superstructure of an ilmenite MnTiO$_3$ including IrO$_6$ honeycomb layers~\cite{MiuraCM,MiuraAPL} could be such a platform, where the antiferromagnetic moment in $\rm MnTiO_3$ can be regarded  as a staggered internal magnetic field acting on the possible Kitaev spin liquid in the IrO$_6$ honeycomb layers~\cite{Haraguchi2018,Haraguchi2020,JM}. 
In such situations, however, the magnon excitations in the antiferromagnet can contribute to the thermal transport. 
In particular, it was shown that the Dzyaloshinskii-Moriya interaction on the second-neighbor bonds leads to nonlinear thermal transport~\cite{HKM,COX,TSM}. 
However, as the Dzyaloshinskii-Moriya interaction arises from the spin-orbit coupling, one may suppress the magnon contribution by choosing the antiferromagnets with less spin-orbit coupling. 
Moreover, we expect that the contribution from Majorana fermions can be distinguished from that from magnons by their temperature dependences, since they obey different statistics. 
Note that this is also the case for phonons, for which we discussed that the field angle dependence is useful in Sec. \ref{subsec:order}. 

\section{Summary}
\label{sec:summary}

To summarize, we have studied the thermal transport in the Kitaev model under uniform and staggered magnetic fields by using the Boltzmann transport theory. 
The staggered field breaks the sublattice symmetry and activates the valley degree of freedom. 
Depending on the amplitudes and directions of the uniform and staggered magnetic fields, the band structure of the Majorana fermions, which are yielded by the fractionalization of spins in the quantum spin liquid state in the Kitaev model, is asymmetrically modulated in momentum space, hosting the gapless or gapped nodal points at finite energy and the Majorana Fermi surfaces. 
We found that the linear thermal current shows a characteristic field dependence, which correlates with the magnitude of the Majorana gap or the effective Majorana density in the presence of the Majorana Fermi surfaces. 
On the other hand, we showed that the asymmetric band modulation gives rise to the nonlinear thermal transport, which also depends on the field amplitude and directions. 
We presented the quantitative estimates by referring to the material parameters of $\a$-RuCl$_3$, and found that the linear thermal response is observable while the improvement of the experimental accuracy is required for the nonlinear component.  
We also discussed how other additional effects of the magnetic fields arising from the second- and third-order perturbation modify the behavior of the thermal transport. 
Furthermore, we discussed contributions from magnons in a heterostructure to realize the internal staggered magnetic field, which would be distinguished from the Majorana contributions.

Our findings illuminate the Majorana band modulations by the uniform and staggered magnetic fields and their observation through the thermal transport measurements, but there remain several open questions. 
The present analysis has been done by the perturbation theory in terms of the magnetic fields, in which the quantum spin liquid state without the localized $Z_2$ fluxes (flux-free state) is retained. 
The flux-free state, however, does not hold beyond the perturbative regime. 
The $Z_2$ fluxes are also excited by raising temperature, which was shown to affect the thermal transport~\cite{NYM}. 
Another interesting unperturbed effect is a field-induced gapless spin liquid state in the antiferromagnetic Kitaev model~\cite{ZKSF,LJCLW,GMP,NKKM,HT,RVT,JDJ}, which would also affect the thermal transport significantly~\cite{TZSSS}. 
In addition, in the candidate materials like $\alpha$-RuCl$_3$, the effect of non-Kitaev interactions, such as the Heisenberg and off-diagonal interactions, cannot be neglected. 
Despite some recent attempts~\cite{YNKM,MKM}, it is still a challenge to access the interesting parameter region at low temperature in applied magnetic fields. 
Further development is required to clarify these important issues.

 \section*{acknowledgement}

We acknowledge S. Okumura, R. Sano, and A. Tsukazaki for fruitful discussions. 
This work is supported by JST CREST (JP-MJCR18T2).

\end{document}